\documentclass[%
superscriptaddress,
bibnotes,
amsmath,amssymb,
aps,
prb, 
longbibliography,
eprint,
floatfix,
reprint
]{revtex4-2}

\usepackage{xcolor}
\definecolor{darkblue}{RGB}{0,0,150}
\definecolor{nightblue}{RGB}{0,0,100}

\usepackage{mathrsfs,dsfont,mathtools,braket}

\newcommand{\refsub}[2]{\hyperref[#1]{\ref*{#1}#2}}

\usepackage{bm}
\usepackage[
colorlinks,
citecolor=darkblue,
linkcolor=darkblue,
urlcolor=nightblue]{hyperref}

\usepackage[english]{babel}
\usepackage[babel,kerning=true,spacing=true]{microtype}

\definecolor{DarkRed}{RGB}{100,0,0}
\definecolor{LightGreen}{RGB}{000,50,0}

\def\({\left(}
\def\){\right)}

\newcommand{\beq}{\begin{equation}}
\newcommand{\eeq}{\end{equation}}
\newcommand{\bal}{\begin{aligned}}
\newcommand{\eal}{\end{aligned}}

\newcommand{\vect}[1]{\boldsymbol{#1}}

\begin{document}

\title{Linear displacement current solely driven by the quantum metric}
\author{Longjun Xiang}
\affiliation{College of Physics and Optoelectronic Engineering, Shenzhen University,
Shenzhen 518060, China}
\author{Bin Wang}
\email{bwang@szu.edu.cn}
\affiliation{College of Physics and Optoelectronic Engineering, Shenzhen University, Shenzhen 518060, China}
\author{Yadong Wei}
\affiliation{College of Physics and Optoelectronic Engineering, Shenzhen University, Shenzhen 518060, China}
\author{Zhenhua Qiao}
\affiliation{Department of Physics, The University of Science and Technology of China, Heifei, P. R. China}
\author{Jian Wang}
\email{jianwang@hku.hk}
\affiliation{College of Physics and Optoelectronic Engineering, Shenzhen University,
Shenzhen 518060, China}
\affiliation{Department of Physics, The University of Hong Kong, Pokfulam Road, Hong Kong, China}
\date{\today}

\begin{abstract}
Quantum metric and Berry curvature are the real part and imaginary part of the quantum geometric tensor,
respectively. The $\mathcal{T}$-odd ($\mathcal{T}$: time-reversal) nonlinear Hall effect driven by
the quantum metric dipole, recently confirmed in [\textcolor{blue}{\textit{Science 381, 181 (2023)}}]
and [\textcolor{blue}{\textit{Nature 621, 487 (2023)}}],
established the geometric duality to the $\mathcal{T}$-even nonlinear Hall effect
that driven by the Berry curvature dipole. Interestingly, a similar geometric duality between the quantum metric
and the Berry curvature, particularly for the \textit{linear response} of Bloch electrons,
has not been established, although the $\mathcal{T}$-odd linear intrinsic anomalous Hall effect (IAHE)
solely driven by the Berry curvature has been known for a long time.
Herein, we develop the quantum theory for displacement current under an AC electric field.
Particularly, we show that the $\mathcal{T}$-even component of the linear displacement current conductivity (LDCC)
is solely determined by the quantum metric, by both the response theory and the semiclassical theory.
Notably, with symmetry analysis we find that the $\mathcal{T}$-even 
LDCC can contribute a \textit{Hall current} in $\mathcal{T}$-invariant systems but with low symmetry,
while its longitudinal component is immune to symmetry. 
Furthermore, employing the Dirac Hamiltonian, we arrive at a $1/\mu$ ($\mu$: chemical potential)
experimental observable enhancement of the displacement current owing to the divergent behavior of quantum metric near Dirac point,
similar to the IAHE at Weyl point. Our work reveals the band geometric origin of the linear displacement current
and establishes, together with the IAHE, the geometric duality for the \textit{linear response} of Bloch electrons.
Additionally, our work offers the very first intrinsic Hall effect in $\mathcal{T}$-invariant materials, 
which can not be envisioned in DC transport in both linear and nonlinear regimes.
\end{abstract}

\maketitle

\section{Introduction}
\begin{figure}[htb!]
\centering
\includegraphics[width=0.85\columnwidth]{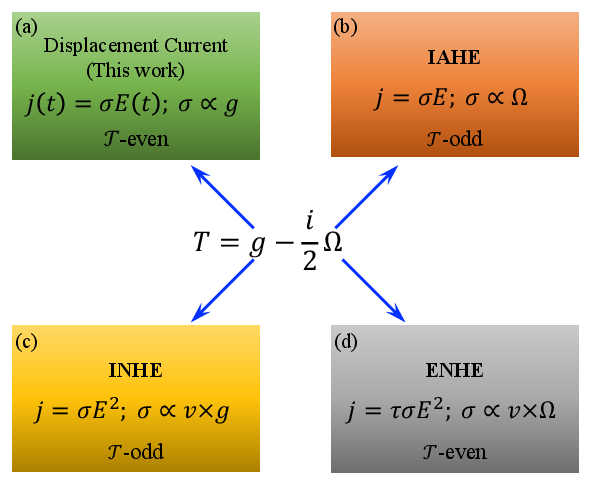}
\caption{
\label{FIG0}
Guided by symmetry, the quantum geometric tensor $T=g-i\Omega/2$ \cite{QGT1} can be fully
probed by four current responses, where $g$ and $\Omega$ stand for the quantum metric
and the Berry curvature, respectively. 
(a) The displacement current probes the $\mathcal{T}$-even quantum metric 
(proposed in this work).
(b) The intrinsic anomalous Hall effect (IAHE) probes the $\mathcal{T}$-odd Berry curvature \cite{Nagaosa2010}.
(c) The intrinsic nonlinear Hall effect (INHE) probes the $\mathcal{T}$-odd
quantum metric dipole $\propto v \times g$ with $v$ the velocity \cite{GaoY2021}.
(d) The extrinsic (proportional to the relaxation time $\tau$)
nonlinear Hall effect (ENHE) probes the
$\mathcal{T}$-even Berry curvature dipole $\propto v \times \Omega$ \cite{FuBCD}.
Note that for the second-order nonlinear responses (c) and (d),
the $\mathcal{P}$-symmetry of the probed system must be broken,
as regulated by the $\mathcal{P}$-odd quantum-metric-dipole and Berry-curvature-dipole.
}
\end{figure}

The quantum geometry \cite{geometry} of Bloch electrons shows a fundamental importance
among various fascinating responses of (topological) quantum materials 
\cite{Nagaosa2017, JEmoore2017, BHYan2021} under electromagnetic fields,
as unveiled by manipulating the symmetry that relates the responses to the band geometric quantities
\cite{light1, light2, light3, symmetry}.
For instance, it has been well understood that the band geometric quantity Berry curvature
is responsible for the intrinsic anomalous Hall effect (IAHE)
observed in ferromagnetic metals \cite{Nagaosa2010},
where the time-reversal ($\mathcal{T}$) symmetry is broken
owing to the IAHE conductivity tensor solely determined by the $\mathcal{T}$-odd Berry curvature \cite{Niu2010},
as illustrated in FIG.(\ref{FIG0}b). Furthermore, 
the Berry curvature dipole \cite{FuBCD, MaQ2019, KK2019, LuNRP2021},
that features the $\mathcal{T}$-even but $\mathcal{P}$-odd ($\mathcal{P}$: space-inversion) properties,
can drive the extrinsic nonlinear Hall effect (ENHE)
in $\mathcal{T}$-invariant but $\mathcal{P}$-broken systems,
as illustrated in FIG.(\ref{FIG0}d).

Apart from the well-known Berry curvature, the quantum geometric tensor of
Bloch electrons also contains a dual band geometric quantity---the quantum metric \cite{QGT1, QGT2, QGT3},
which recently received much attention,
but mainly in the form of quantum metric dipole that usually appears in the nonlinear transport of Bloch electrons
\cite{GaoY2014, GaoY2021, XiaoC2021, XuSY2023, Gao2023, CulcerQMD, ZGZhuQMD}
and features the $\mathcal{T}$-odd and $\mathcal{P}$-odd properties.
Among them, the quantum metric dipole driven intrinsic nonlinear Hall effect (INHE) \cite{GaoY2014, GaoY2021, XiaoC2021},
as illustrated in FIG.(\ref{FIG0}c),
has been recently confirmed experimentally in antiferromagnetic topological insulator MnBi$_2$Te$_4$ \cite{XuSY2023, Gao2023}
(that breaks $\mathcal{P}$ and $\mathcal{T}$ symmetries).
However, how the quantum metric itself is manifested in the \textit{linear response} of Bloch electrons
(as the dual effect of the IAHE) remains elusive, in sharp contrast with the Berry curvature.

On the other hand, we notice that the responses discussed above,
as the nonlinear variants of the Ohm's law \cite{ZhangScience},
generally feature the Fermi-surface property \cite{FuBCD, GaoY2014, GaoY2021, XiaoC2021}
and thereby can only be expected in metallic or gapless systems.
However, in addition to the conduction current that appears in the metallic solids,
the current in Maxwell's equations \cite{Jackson} in fact contains another contribution,
namely the displacement current that appears in the insulating solids \cite{Marder}.
Although the connection between the linear/nonlinear responses
and the band geometric quantities of the conducting Bloch electrons (electrons on the Fermi-surface),
has been well established, how the displacement current (even at linear regime)
relates to the band geometric quantity is unexplored.
In particular, as the cousin of the conduction current, 
whether or not there exists a Hall effect for the displacement current is also unknown.

In this work, inspired by the intimate relation between the electric polarization and the displacement current,
we develop the quantum theory for the linear displacement current under an AC electric field based on
the quantum response theory.
We show that the linear displacement current conductivity (LDCC)
comprises both the $\mathcal{T}$-even and the $\mathcal{T}$-odd contributions.
Particularly, we reveal that the $\mathcal{T}$-even DCC is solely
determined by the quantum metric.
Remarkably, by symmetry analysis, we find that the $\mathcal{T}$-even DCC can
allow a transverse component and hence displays a
\textit{displacement Hall current} in $\mathcal{T}$-invariant systems but with low symmetry,
while its longitudinal component is less restricted by symmetry.
As a benchmark, we show that the obtained expressions in adiabatic limit
can also be derived by the semiclassical theory accurate up to the second order.
Furthermore, using the low-energy Dirac Hamiltonians,
we find that the displacement current will be significantly enhanced when 
the chemical potential is close to the Dirac point where the quantum metric is divergent,
similar to the behavior of IAHE at Weyl point.
Our work uncovers the band geometric origin of the linear displacement current
and establishes, together with the IAHE, the geometric duality for the \textit{linear response} of Bloch electrons,
as illustrated in FIG.(\ref{FIG0}a-\ref{FIG0}b).
In addition, our work delivers the very first intrinsic Hall effect in $\mathcal{T}$-invariant systems,
which can not appear in DC transport in both linear and nonlinear regimes,
particularly due to the $\mathcal{T}$-odd nature of the intrinsic DC linear
and nonlinear conductivities.

\bigskip
\section{Displacement current from electric polarization}
Before presenting the quantum theory,
it is instructive to give a heuristic argument on the origin of displacement current.
In terms of Maxwell's classical electromagnetic theory,
for insulating or gapped systems, the conduction current vanishes due to the lack of free charges.
However, an AC electric field $E_\beta(t)=E_\beta\cos(\omega t)$ can generate
a displacement current density $J^D_\beta=\partial_t D_\beta$,
where $D_\beta \equiv E_\beta (t) + 4 \pi \epsilon_0 P_\beta (t)$ is the displacement field
with $\epsilon_0$ and $P_\beta$ the vacuum dielectric constant and
the electric polarization induced by the positional shift of bound charges \cite{Nagaosa}, respectively.
Interestingly, although the (spontaneous) electric polarization of crystalline solids
has been well understood through the geometric phase of Bloch electrons \cite{Niu2010},
the displacement current seems not to appreciate any geometrical effects and also Hall effect, as far as we know.

Up to the first order of electric field,
we find that $P_\beta(t)$ can be expressed as 
\cite{Niu2010, KS-V1993, V-KS1993, Resta1994, XiaoCmagnetization} ($\hbar=e=1$),
\begin{align}
P_\beta (t)
=
-\sum_n \int_k f_n
\left[ \mathcal{A}_{n}^{\beta} + \mathcal{A}_{n}^{\beta,E} (t) \right],
\label{polar}
\end{align}
where $\int_k \equiv \int dk^d/(2\pi)^d$ with $d$ the dimension of the system and
$f_n$ is the equilibrium Fermi distribution function.
In Eq.(\ref{polar}), the intraband Berry connection $\mathcal{A}_n^\beta$
contributes the spontaneous or zero-field electric polarization
\cite{KS-V1993, V-KS1993, Resta1994, XiaoCmagnetization},
while $\mathcal{A}_n^{\beta,E}(t)\equiv G^{\beta\alpha}_nE_\alpha(t) $
is responsible for the field-induced electric polarization linear in the electric field
\cite{Sipe1995, Gonze2001, Vanderbilt2002}.
In particular, $G^{\beta\alpha}_n$ is the Berry connection polarizability tensor
given by \cite{XuSY2023, Gao2023, GaoY2014, YSA2021, Wang2022}
\begin{align}
G^{\beta\alpha}_n
\equiv
2\text{Re} \sum_{m}
\dfrac{r^\beta_{nm} r^{\alpha}_{mn}}{\epsilon_{n}-\epsilon_{m}},
\label{staticGab}
\end{align}
where $r^\beta_{nm} \equiv \langle u_n|i\partial_\beta |u_m\rangle$ with $m \neq n$ ($|u_n\rangle$: the periodic part of Bloch state)
is the interband Berry connection
and $\epsilon_n$ is the energy for the $n$th band.
We wish to mention that $G^{\beta\alpha}_n$ encodes the information of local quantum metric
$g^{\beta\alpha}_{nm}=2 \text{Re} [r^{\beta}_{nm}r^{\alpha}_{mn}]$.
Particularly, both quantities possess the same symmetry transformation.

By definition, taking the time derivative of field-induced electric polarization,
we obtain a linear displacement current up to the first order in the driving frequency
with a response equation
\begin{eqnarray}
J^D_\beta = \sigma^{D}_{\beta\alpha}(t) E_\alpha,
\label{DHEresponserelation}
\end{eqnarray}
where
\begin{eqnarray}
\sigma^{D}_{\beta\alpha}(t) = \omega \sin(\omega t) \sum_n \int_k f_n G^{\beta\alpha}_n
\label{first}
\end{eqnarray}
is the LDCC. Note that Eq.(\ref{DHEresponserelation}) displays a formal similarity to the IAHE
since they are driven by the real part and the imaginary part
of the quantum geometric tensor, respectively, as compared in FIG.(\ref{FIG0}a) and FIG.(\ref{FIG0}b).
In fact, they can be derived in a unified way, as will be shown below.
Interestingly, we find that $\sigma_{\beta\alpha}^D$
may include a transverse component $\sigma^D_{yx}$
when the integral of $G^{yx}_n$ does not vanish.
Therefore, a displacement Hall effect in $\mathcal{T}$-invariant systems
can be expected due to $\mathcal{T} G^{\beta\alpha}_n=G^{\beta\alpha}_n$.

We wish to remark that the intrinsic linear (charge) Hall effect is forbidden by $\mathcal{T}$-symmetry in the DC case,
as dictated by Onsager relation \cite{Nagaosa2010} and the $\mathcal{T}$-odd nature for
the corresponding response coefficient. In general, the linear DC Hall conductivity can be decomposed as
$\sigma_{\alpha\beta} = \sigma_{\alpha\beta}^{\text{in}} + \tau \sigma_{\alpha\beta}^{\text{ex}}$ \cite{XiaoCspin, XiaoCdynamical},
where the first term is intrinsic but $\mathcal{T}$-odd
while the second term is $\mathcal{T}$-even but extrinsic
due to the presence of the relaxation time $\tau$ \cite{XiaoC2023}.
However, for displacement current, we find that the time derivative for
the field-induced electric polarization replaces the role of $\tau$ in DC case \cite{XiaoCdynamical}
and hence allows the intrinsic $\mathcal{T}$-even linear Hall effect for the displacement current,
which does not show up in DC charge transport.

Up to now, we only present a qualitative discussion.
Particularly, Eq.(\ref{first}) is a hand waving derivation and
the rigorous one should be obtained with quantum mechanical calculations, as detailed below.
Additionally, we wish to remark that Eq.(\ref{first}) seems not applicable to gapless systems
where the concept of electric polarization is not well defined \cite{XiaoCspin} 
particularly due to the static electric screening.
However, this constraint can be relaxed in AC transport with a high frequency \cite{screen} .

\begin{figure*}[t!]
\centering
\includegraphics[width=1.80\columnwidth]{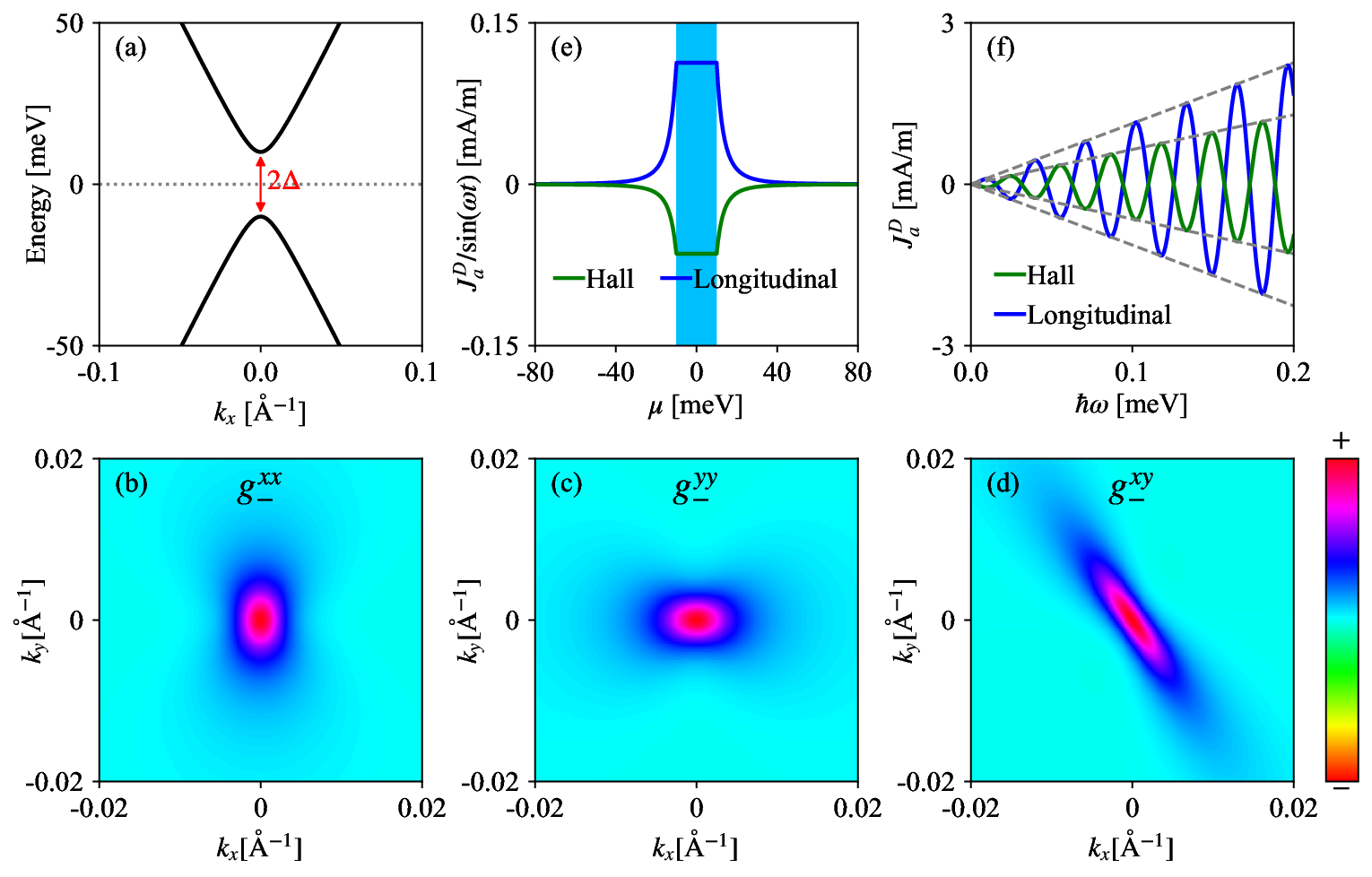}
\caption{
\label{FIG1}
(a) The band dispersion for Eq.(\ref{Ham1}) along $k_x$ direction.
The red vertical arrow measures the gap.
The $\vect{k}$-resolved distribution for the quantum metric (a) $g^{xx}_{-}$,
(b) $g^{yy}_{-}$, and (c) $g^{xy}_{-}$ [Here a term $sv_yk_x\sigma_x$ is added to break the mirror symmetries of Eq.(\ref{Ham1})].
(e) The $\mu$ dependence for the longitudinal and Hall displacement current,
given by Eq.(\ref{JxJy}) and Eq.(\ref{JxJy1}), respectively,
by choosing the driving frequency $\omega=0.01 (\mathrm{meV})$.
Here the vertical shadow highlights the in-gap current plateau.
(f) The frequency dependence for the longitudinal and Hall displacement current
when $\mu$ is placed inside the gap. Here the dashed grey line indicates the linear dependence.
}
\end{figure*}

\bigskip
\section{The quantum theory for displacement current}
In this section, we formulate the quantum theory for displacement current
under an AC electric field based on the quantum response theory \cite{Sipe2000, JEMoore2010, JustingSong}.
To be specific, we start from the quantum Liouville equation
for the density matrix element $\rho_{mn}$ \cite{Sipe2000},
\begin{align}
i\dfrac{\partial}{\partial t} \rho_{mn}
&=
\epsilon_{mn}\rho_{mn}+ i(\rho_{mn})_{;k_\alpha} E_\alpha (t) \nonumber \\
&+
\sum_{l}\left[r_{ml}^\alpha \rho_{ln}(t)- \rho_{ml}(t) r_{ln}^\alpha \right]E_\alpha (t),
\label{rhomnequation}
\end{align}
where $(\rho_{mn})_{;k_\alpha}= \left[\partial_\alpha -i (\mathcal{A}^\alpha_m -\mathcal{A}^\alpha_n) \right] \rho_{mn}$
with $\partial_\alpha \equiv \partial/\partial k_\alpha$,
$r_{ml}^\alpha$ is the interband Berry connection and
$E_\alpha(t)=E_\alpha \cos(\omega t)=\frac{1}{2}\sum_{\omega_1} E_\alpha e^{i\omega_1 t}$
with $\omega_1=\pm \omega$ is the AC electric field. Eq.(\ref{rhomnequation}) can be recursively solved to obtain a series
in increasing powers of the electric field, namely $\rho_{mn}=\sum_{n=1}^\infty \rho_{mn}^{(n-1)}$,
where $\rho_{mn}^{(n-1)}$ is proportional to $E_\alpha^{(n-1)}$.
To our purpose, with $\rho^{(0)}_{nm}=\delta_{nm} f_n$ \cite{Sipe2000},
the density matrix $\rho^{(1)}_{nm}$ at the first order of electric field
can be easily obtained from Eq. (\ref{rhomnequation}), which is
\begin{align}
\rho_{mn}^{(1)}
=
\dfrac{1}{2}
\sum_{\omega_1}
\left[
i\delta_{mn}\partial_{\alpha}f_m
+
f_{nm}r^\alpha_{mn}
\right]
\dfrac{E_\alpha e^{-i(\omega_1+i\eta) t}}{\omega_1-\epsilon_{mn}+i\eta},
\end{align}
where $f_{nm}=f_n-f_m$ and $\eta \rightarrow 0^{+}$ is an infinitesimal quantity.
Correspondingly, the current density at the first order of the electric field,
defined as $J_\beta \equiv \sum_{mn} \int_k v^\beta_{nm} \rho^{(1)}_{mn}$, is found to be
\begin{align}
J_\beta
&=
\sum_{n} \sum_{\omega_1}
\int_k v^\beta_{n}
\partial_{\alpha} f_{n} E_\alpha \dfrac{i e^{-i(\omega_1+i\eta)t}}{2(\omega_1+i\eta)}
\nonumber\\
&+\sum_{mn} \sum_{\omega_1}  \int_k
f_{nm} r^\beta_{nm} r^\alpha_{mn}
E_\alpha
\dfrac{i\epsilon_{nm} e^{-i (\omega_1+i\eta)t}}{2(\omega_1-\epsilon_{mn}+i\eta)},
\label{response}
\end{align}
where we have used $v^\beta_{nm}=i\epsilon_{nm}r^\beta_{nm}$ for $n \neq m$.
Note that the first term of Eq.(\ref{response})
is irrelevant to the band geometry of Bloch electrons
and corresponds to the familiar extrinsic linear Drude current in DC limit 
due to $\lim_{\omega_1 \rightarrow 0} i/(\omega_1+i\eta) = \tau$ \cite{FuBCD}.
Besides the extrinsic linear Drude current, we note that
Eq.(\ref{response}) (particularly by the nonresonant contribution of its second term)
further contains the well-known intrinsic anomalous Hall current and
the overlooked linear displacement current phenomenally discussed above,
which is explicitly expressed as
\begin{align}
J_\beta^N
=
\sum_{mn} \sum_{\omega_1}  \int_k
f_{nm} r^\beta_{nm} r^\alpha_{mn}
E_\alpha
\dfrac{i\epsilon_{nm} e^{-i \omega_1 t}}{2(\omega_1-\epsilon_{mn})}.
\label{ACDC}
\end{align}
At this stage, by writing $\epsilon_{nm}/(\omega_1-\epsilon_{mn})=1-\omega_1/(\omega_1-\epsilon_{mn})$,
we find that $J_\beta$ can be divided into (see Appendix \ref{AppA})
\begin{align}
J_\beta^N = J^C_\beta + J^D_\beta \label{CandD},
\end{align}
where $J^C$ is the intrinsic AC conduction Hall current, given by
\begin{align}
J^C_\beta &= \sum_{n}\int_k f_n\Omega^{\beta\alpha}_n E_\alpha \cos\omega t \label{hall1},
\end{align}
where $\Omega_{n}^{\beta\alpha}=2 \sum_{m} \text{Im} [r^{\beta}_{nm}r^{\alpha}_{mn}]$
is the Berry curvature. In DC limit,
Eq.(\ref{hall1}) is nothing but the intrinsic anomalous Hall current.
Furthermore, $J^D_\beta = \partial_t P_\beta$ stands for the intrinsic displacement current,
where $P_\beta$ is the AC polarization
\begin{align}
P_\beta
&=
\sum_{n}\int_k f_{n}
\left[
\mathcal{G}^{\beta\alpha}_n \cos\omega t
+
\mathcal{F}^{\beta\alpha}_n \sin\omega t
\right]
E_\alpha.
\label{ACP}
\end{align}
Notably, $\mathcal{G}^{\alpha\beta}_n$ and $\mathcal{F}^{\alpha\beta}_n$ in Eq.(\ref{ACP}) 
encode the information of quantum metric and Berry curvature, respectively, given by
\begin{align}
\mathcal{G}^{\beta\alpha}_n
&=
\sum_m 
\dfrac{\epsilon_{mn}}{\omega^2- \epsilon_{mn}^2}
2\text{Re} \left[r^\beta_{nm} r^\alpha_{mn}\right],
\label{Gab} \\
\mathcal{F}^{\beta\alpha}_n
&=
\sum_m 
\dfrac{\omega}{\omega^2- \epsilon_{mn}^2}
2\text{Im} \left[ r^\beta_{nm} r^\alpha_{mn} \right].
\label{Fab}
\end{align}
Under $\mathcal{T}$-symmetry, it is easy to show that
$\mathcal{T} \mathcal{G}^{\alpha\beta}_{n}=\mathcal{G}^{\alpha\beta}_n$
while $\mathcal{T} \mathcal{F}^{\alpha\beta}_{n}=-\mathcal{F}^{\alpha\beta}_n$
due to $\mathcal{T} r^\alpha_{nm}=r^\alpha_{mn}$,
and therefore the displacement current in general includes both
the $\mathcal{T}$-even and $\mathcal{T}$-odd contributions,
similar to the DC shift and injection photocurrent under light illumination \cite{Sipe2000, Wanghua2020}.
However, $\mathcal{T}$-even component becomes dominant in the adiabatic limit (the first order in the driving frequency $\omega$)
and thereby we will focus on this contribution in the following. 
In that limit, we find that the AC polarization $P_\beta$ recovers the field-induced polarization
[the second term of Eq.(\ref{polar})].

As expected, we find that the displacement current features a Fermi-sea form,
which means that the expression for $J^D_\beta$ can be used for the insulating and also
the metallic systems. Although the intrinsic displacement current particularly contributed by Eq.(\ref{Gab})
is formally similar to Eq.(\ref{hall1}), they indeed behaves differently. 
Firstly, for topological trivial/nontrivial insulators, Eq.(\ref{hall1}) gives a zero/quantized conductivity;
while for magnetic metals, Eq.(\ref{hall1}) will give the IAHE
contributed by the electrons on the Fermi surface \cite{Niu2010}. 
Secondly, the displacement current vanishes in the DC limit ($\omega \rightarrow 0$)
while Eq.(\ref{hall1}) can survive in that limit.
Thirdly, Eq.(\ref{hall1}) probes the Berry curvature while the displacement current probes the quantum metric,
as compared in FIG.(\ref{FIG0}a) and FIG.(\ref{FIG0}b).
To close this section, we summarize that Eqs.(\ref{ACP}-\ref{Gab}) are the main results in this work,
which can also be derived by the semiclassical theory in adiabatic limit \cite{Niu2010},
see Appendix \ref{AppB}.

\bigskip
\section{Symmetry constraints}
After establishing the quantum theory for the displacement current,
we now discuss the symmetry constraints for its conductivity,
particularly on its $\mathcal{T}$-even component from 32 crystallographic point groups.
Note that the number of independent components of a physical tensor
is determined by Neumann's principle \cite{anisotropy}.
Particularly, for the rank-2 LDCC tensor, we have
\begin{align}
\sigma^{D}_{\beta\alpha}= R_{\beta\beta'} R_{\alpha\alpha'}
\sigma^{D}_{\beta'\alpha'},
\label{symmetry}
\end{align}
where $R_{\beta\beta'}$ stands for the matrix element for
the point group operation $R$.
Note that Eq.(\ref{symmetry}) has been implemented by Bilbao Crystallographic Server \cite{bilbao}.
As a consequence, by defining the Jahn notation $[V^2]$ \cite{KTlaw} for $\sigma^D_{\beta\alpha}$,
we find that $\sigma_{\beta\beta}^D$ is allowed by all 32 crystallographic point groups
while $\sigma_{\beta\alpha}^D$ with $\alpha \neq \beta$
can only appears in crystallographic point groups with low symmetry:
$1, -1, 2, m, 2/m$. Here we wish to remark that
the transverse components of $\sigma^D_{\beta\alpha}$
offers the very first intrinsic Hall effect in time-reversal-invariant systems,
which can not appear in DC transport since all the intrinsic DC transport coefficients
feature the $\mathcal{T}$-odd nature.

\begin{figure*}[t!]
\centering
\includegraphics[width=1.85\columnwidth]{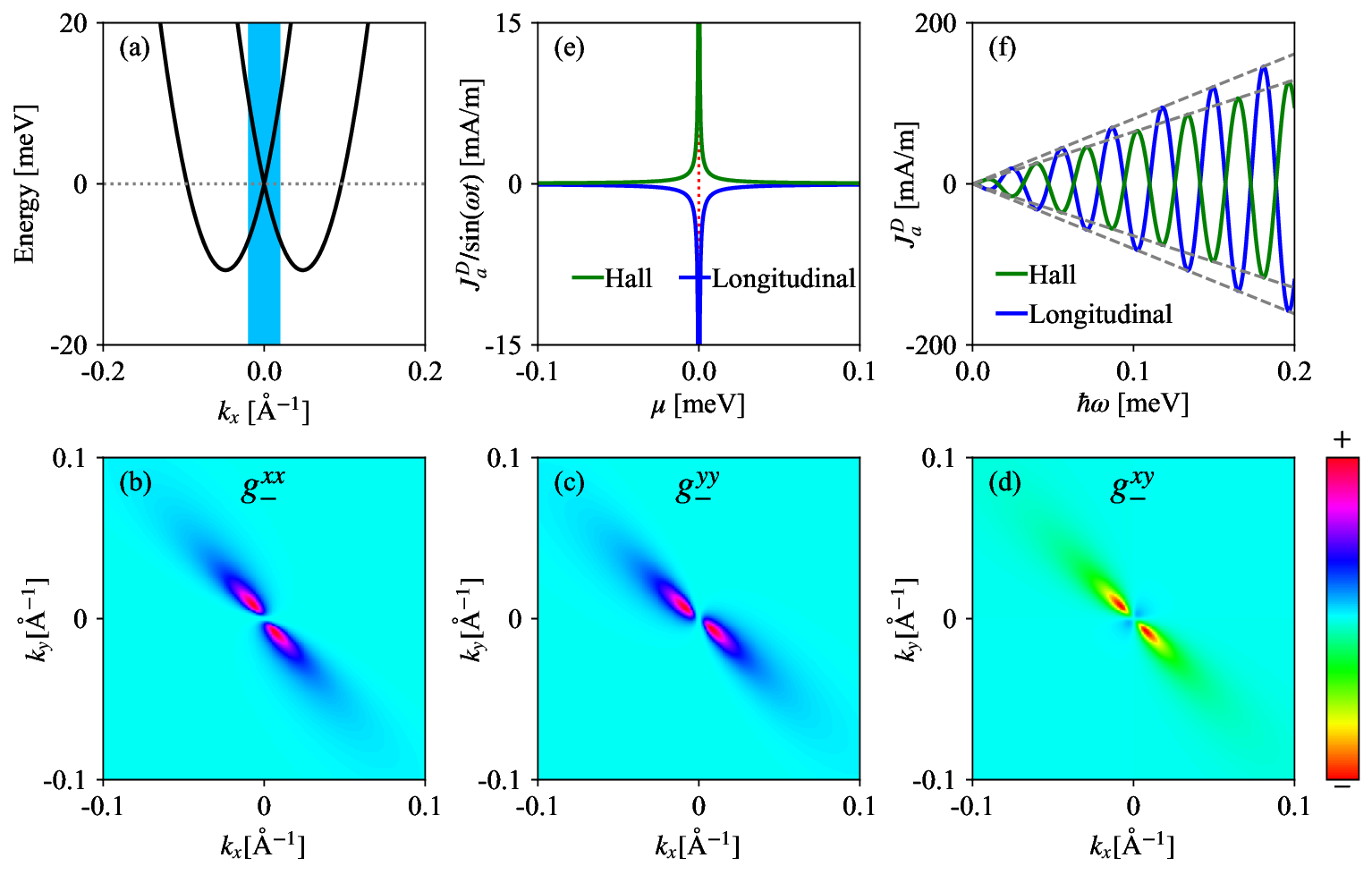}
\caption{
\label{FIG2}
(a) The band dispersion for Eq.(\ref{HRD}) along $k_x$ direction. Note that the vertical shadow highlights regime that the displacement current will dominate.
The $\vect{k}$-resolved distribution for the quantum metric (a) $g^{xx}_{-}$,
(b) $g^{yy}_{-}$, and (c) $g^{xy}_{-}$.
(e) The $\mu$ dependence for the longitudinal and Hall displacement current,
given by Eq.(\ref{sigmaD}),
by choosing the driving frequency $\hbar\omega=10^{-4} (\mathrm{meV})$.
Here the dashed red line highlights the divergent behavior at $\mu=0$.
(f) The frequency dependence for the longitudinal and Hall displacement current
when $\mu=0.1 \mathrm{meV}$. 
}
\end{figure*}

\bigskip
\section{Model Calculations}
In this section, we illustrate our theory with model Hamiltonians.

\bigskip
\subsection{Two-dimensional massive Dirac model}
In this section, we employ a two-band massive model to illustrate our general proposal.
The low energy Hamiltonian to describe the massive Dirac fermions located at momenta $\Lambda$
is given by \cite{FuBCD}
\begin{align}
H_{s\Lambda}=v_xk_x \sigma_y - s v_y k_y \sigma_x + \Delta \sigma_z,
\label{Ham1}
\end{align}
where $\sigma_i$ is the Pauli matrix for pseudospin, $s = \pm 1$, 
$\Delta$ controls the gap magnitude, and $v_{x/y}$ is the Fermi velocity.
For simplicity, we ignore the tilt term $s \alpha k_y$ in Ref.[\onlinecite{FuBCD}].
The band dispersion for Eq.(\ref{Ham1})
is $\epsilon_{\pm}= \pm h$ with $h=\sqrt{\Delta^2+v_x^2 k_x^2 + v_y^2k_y^2}$,
where $+$ $(-)$ denotes the conduction (valence) band, as shown in FIG.(\ref{FIG1}a).
Note that Eq.(\ref{Ham1}) possesses the mirror symmetry $\mathcal{M}_y$ due to $\mathcal{M}_y s=-s$,
which indicates that the transverse LDCC vanishes by Eq.(\ref{symmetry}).
For the longitudinal LDCC, using Eq.(\ref{Gab}) we find
\begin{align}
\mathcal{G}_{\pm}^{xx} &= \pm \dfrac{v_x^2\left(v_y^2k_y^2+\beta^2\right)}{4h^3(h^2-\omega^2)},
\\
\mathcal{G}_{\pm}^{yy} &= \pm \dfrac{v_y^2\left(v_x^2k_x^2+\beta^2\right)}{4h^3(h^2-\omega^2)}.
\end{align}
It is easy to see that both $\mathcal{G}^{xx}_{\pm}$ and $\mathcal{G}^{yy}_{\pm}$ that even in $k_x$ and $k_y$
can lead a nonvanishing longitudinal current.
We emphasize that $\mathcal{G}^{xx}_{\pm}$ and $\mathcal{G}^{yy}_{\pm}$
encode the diagonal contribution of the quantum metric, as displayed in FIG.(\ref{FIG1}b) and FIG.(\ref{FIG1}c), respectively.
At zero temperature, the longitudinal displacement current can be directly calculated as (see Appendix \ref{AppC1})
\begin{align}
\dfrac{J^D_{x}}{E_x}
=
\dfrac{J^D_{y}}{E_y}
=
\begin{cases}
e^2 \dfrac{\Delta^2 7 \omega \sin(\omega t)}{48 \pi |\mu|^3} \quad \mu \notin [-\Delta, \Delta], \\[0.25cm]
e^2 \dfrac{7 \omega \sin(\omega t)}{48 \pi \Delta} \quad \mu \in [-\Delta, \Delta],
\end{cases}
\label{JxJy}
\end{align}
where $\mu$ is the chemical potential.
Note that have used $\Delta \gg \hbar\omega$ and $v_x=v_y$ \cite{FuBCD}.
Here the elementary charge $e$ is restored by dimension analysis.
By Eq.(\ref{JxJy}), we find that the longitudinal displacement current will
quickly decrease to zero when the chemical potential $\mu$ touches to the bands,
whereas a longitudinal LDCC plateau can be obtained when $\mu$ is located inside the gap,
as shown in FIG.(\ref{FIG1}e). Note the the longitudinal displacement current shows
a linear dependence on the driving frequency $\omega$, as shown in FIG.(\ref{FIG1}f) and therefore will
disappear in DC transport. For the gapped situation,
one can easily obtain the AC polarization from the displacement current, 
$P_{x}/E_x
=
P_{y}/E_y
=
7 e^2/(48 \pi \Delta) \cos(\omega t)
$,
which shows explicitly that AC polarization will be stronger when the gap decreases.

To acquire a nonzero displacement Hall current,
we break the mirror symmetry $\mathcal{M}_y$ of Eq.(\ref{Ham1}) by introducing an additional term $sv_yk_x\sigma_x$,
which may be physically realized by strain effects \cite{strain}, and then we have
\begin{align}
G^{xy}_{\pm}
=
\mp \dfrac{v_y^2(v_x^2k_xk_y-\Delta^2)}{4(v_x^2k_x^2+v_y^2(k_x+k_y)^2+\Delta^2)^{5/2}},
\label{Gxy}
\end{align}
where we have adopted the adiabatic limit. Eq.(\ref{Gxy})
encodes the off-diagonal information of the quantum metric, as shown in FIG.(\ref{FIG1}d).
In a similar way, the displacement Hall current at zero temperature is found to be (see Appendix \ref{AppC2})
\begin{align}
\dfrac{J^D_{x}}{E_y}
=
\begin{cases}
- e^2 \dfrac{\Delta^2 \omega \sin(\omega t)}{12 \pi |\mu|^3} \quad \mu \notin [-\Delta, \Delta], \\[0.25cm]
- e^2 \dfrac{\omega \sin(\omega t)}{12 \pi \Delta} \quad \mu \in [-\Delta, \Delta],
\end{cases}
\label{JxJy1}
\end{align}
which shows the same behavior with the longitudinal displacement current,
as compared in FIG.(\ref{FIG1}e) and FIG.(\ref{FIG1}f). Note that the magnitude of the
linear displacement current is determined by the dimensionless factor $\sim \hbar\omega/\Delta$,
which indicates that this effect will be significantly enhanced when the gap $\Delta \rightarrow 0$
when the driving frequency $\omega$ is fixed, as shown below.

\bigskip
\subsection{Two-dimensional spin-orbit-coupled electron gas}
Next we evaluate the displacement current with a gapless model.
Particularly, we consider the low-energy effective Hamiltonian 
for two-dimensional spin-orbit-coupled electron gas with crystallographic point group $2$,
which is constructed by adding Rashba and Dressellhaus spin-orbit coupling and is written as
\cite{RD1, RD2, RD3}
\begin{align}
H=
E_0
+
\lambda_R(k_y\sigma_x-k_x\sigma_y)
+
\lambda_D(k_x\sigma_x-k_y\sigma_y),
\label{HRD}
\end{align}
where $E_0=k^2/2m$ with $k^2=k_x^2+k_y^2$.
The band dispersions for Eq.(\ref{HRD}) are given by
$\epsilon_{\pm}=E_0 \pm \sqrt{(\lambda_D k_y+\lambda_R k_x)^2+(\lambda_D k_x+\lambda_R k_y)^2}$
with $\pm$ the valance band and conduction band, respectively, as shown in FIG.(\ref{FIG2}a).
Focusing on the band crossing regime (as highlighted by the vertical shadow in FIG.(\ref{FIG2}a)),
which in fact corresponds to the massless Dirac Hamiltonian,
the linear displacement Hall current under adiabatic limit at zero temperature can be evaluated (see Appendix \ref{AppC3}):
\begin{align}
J_a^D
=
\dfrac{e^2}{\hbar} \dfrac{\hbar \omega }{|\mu|} C_{ab} E_b \sin(\omega t) \quad a, b \in \{x,y\},
\label{sigmaD}
\end{align}
where $C_{xy}=-\lambda_R \lambda_D/[8\pi(\lambda_R^2-\lambda_D^2)]$ and 
$C_{xx}=C_{yy}=(\lambda_R^2 + \lambda_D^2)/[16\pi(\lambda_R^2-\lambda_D^2)]$.
Note that Eq.(\ref{sigmaD}) shows a $|\mu|^{-1}$ dependence and
doesn't show a conductivity plateau due to the lack of a finite gap, as shown
FIG.(\ref{FIG2}e). Due to this divergent behavior, an enhanced displacement current can be achieved with
a relatively low driving frequency compared to the gapped situation, as shown in FIG.(\ref{FIG2}f). In addition,
in FIG.(\ref{FIG2}b-\ref{FIG2}d), the corresponding $\vect{k}$-resolved distribution
for the diagonal and off-diagonal components of the quantum metric are shown,
which are responsible for the divergent behavior of the linear displacement current,
in a similar way to the divergent IAHE in the point-node Weyl point.

\bigskip
\section{Summary}
In summary, we formulate the quantum theory for displacement current under an AC electric field.
We show that the $\mathcal{T}$-even LDCC is solely determined by the quantum metric.
In terms of symmetry analysis, we find that the longitudinal component of the LDCC is immune
to symmetry meanwhile its Hall component can be expected even in $\mathcal{T}$-invariant systems
but with low symmetry, such as the strained transition metal dichalcogenides (TMDCs) monolayers,
the strained surface of topological crystalline insulators,
and the insulating two-dimensional chiral twisting graphene or TMDCs \cite{twist1, twist2, twist3, twist4, twist5}.
Furthermore, our model calculations demonstrate that an enhanced displacement current
can be achieved in the massless Dirac point, such as in two-dimensional graphene monolayer,
where the quantum metric shows a divergent behavior like the Berry curvature in the Weyl point.
Given the fact that the experimental measurement on the ENHE due to the Berry curvature dipole
is already in the AC regime with small frequency to lock the nonlinear conductivity,
the LDCC can be measured experimentally.
Our work reveals the band geometric origin of the linear displacement current
and in turn, offers a desirable tool to detect the quantum metric of Bloch electrons in quantum materials.
Importantly, our work together with the IAHE establishes the geometric duality
for the \textit{linear response} of Bloch electrons. 
In addition, the transverse displacement current also provides the very first Hall effect
in $\mathcal{T}$-invariant materials, which is forbidden by $\mathcal{T}$-symmetry
in DC transport.
Beyond these, the linear displacement current
in point-node Weyl semimetals (which also appreciates the $1/\mu$ divergent behavior)
may be employed to realize the low-energy high-speed photodetection particularly in terahertz regime
\cite{JAhn, terahertz, terahertz1}, which will be explored in the future.

\smallskip
\section*{Acknowledgements}
J. W. thanks the financial support from the National Natural Science Foundation of China (Grant No. 12034014).

\bigskip
\onecolumngrid
\appendix

\section{The detailed derivations for Eqs.(\ref{hall1}-\ref{Fab})}
\label{AppA}
In this appendix, we show the detailed derivation for Eqs.(\ref{hall1}-\ref{Fab}) given in the main text.
Particularly, by writing 
\begin{align}
\dfrac{\epsilon_{nm}}{\omega_1-\epsilon_{mn}}=1-\dfrac{\omega_1}{\omega_1-\epsilon_{mn}},
\end{align}
we find that $J_\beta^N$ given by Eq.(\ref{ACDC}) in the main text can be divided into
\begin{align}
J_\beta^N &= J^C_\beta + J^D_\beta,
\end{align}
where 
\begin{align}
J^C_\beta 
\equiv
\dfrac{1}{2}\sum_{nm} \sum_{\omega_1=\pm\omega} \int_k f_{nm} r^\beta_{nm} r^\alpha_{mn} E_\alpha i e^{-i\omega_1 t}
=
\sum_{nm} \int_k f_{nm} r^\beta_{nm} r^\alpha_{mn} E_\alpha i \cos(\omega t)
=
\sum_{n}\int_k f_n\Omega^{\beta\alpha}_n E_\alpha \cos\omega t,
\end{align}
as given by Eq.(\ref{hall1}) in the main text. Here the last term
is obtained by interchanging the dummy index
and $\Omega_{n}^{\beta\alpha}=i\sum_m(r^\beta_{nm}r^\alpha_{mn}-r^\alpha_{nm}r^\beta_{mn})$ is the Berry curvature.
Furthermore, for the remaining term of $J_\beta^N$, we have:
\begin{align}
J_\beta^D
&=
\sum_{nm} \sum_{\omega_1}
f_{nm} r^\beta_{nm} r^\alpha_{mn} E_\alpha \dfrac{-i\omega_1e^{-i\omega_1t}}{2(\omega_1-\epsilon_{mn})}
=
\partial_t P_\beta(t),
\end{align}
where
\begin{align}
P_\beta 
&\equiv 
\sum_{nm} \sum_{\omega_1}
f_{nm} r^\beta_{nm} r^\alpha_{mn} E_\alpha \dfrac{e^{-i\omega_1t}}{2(\omega_1-\epsilon_{mn})}
=
\sum_{nm}
f_{nm} r^\beta_{nm} r^\alpha_{mn} E_\alpha
\left[
\dfrac{e^{i\omega t}}{2(-\omega-\epsilon_{mn})}
+
\dfrac{e^{-i\omega t}}{2(\omega-\epsilon_{mn})}
\right]
\nonumber \\
&=
\sum_{nm}
f_{nm} r^\beta_{nm} r^\alpha_{mn} E_\alpha
\dfrac{\epsilon_{mn}\cos(\omega t)-i \omega \sin(\omega t)}{\omega^2-\epsilon_{mn}^2}
=\sum_{n}\int_k f_{n} 
\left[
\mathcal{G}^{\beta\alpha}_n \cos\omega t
+
\mathcal{F}^{\beta\alpha}_n \sin\omega t
\right]
E_\alpha,
\label{ACPappendix}
\end{align}
as given by Eq.(\ref{ACP}) in the main text.
Note that the last term of Eq.(\ref{ACPappendix}) is also obtained by interchanging dummy index and
$\mathcal{G}^{\alpha\beta}_n$ and $\mathcal{F}^{\alpha\beta}_n$, respectively,
is defined as
\begin{align}
\mathcal{G}^{\beta\alpha}_n
&= 
\sum_m \dfrac{\epsilon_{mn}}{\omega^2- \epsilon_{mn}^2}
2\text{Re} \left[r^\beta_{nm} r^\alpha_{mn}\right],
\label{Gabappendix} \\
\mathcal{F}^{\beta\alpha}_n
&= 
\sum_m \dfrac{\omega}{\omega^2- \epsilon_{mn}^2} 
2 \text{Im} \left[ r^\beta_{nm} r^\alpha_{mn} \right],
\label{Fabappendix}
\end{align}
as given by Eq.(\ref{Gab}) and Eq.(\ref{Fab}) in the main text.

\section{The semiclassical formulation for displacement current}
\label{AppB}
In this appendix, we show that the displacement current can also be derived with the semiclassical theory
accurate up to the second order of the electric field.
In particular, under a time-dependent but slowly perturbed electric field,
the semiclassical equation of motions for Bloch electrons
up to the second order of electric field are given by \cite{GaoY2014, Niu2010, XiaoCmagnetization, Y-Gao1, Qiao}
\begin{align}
\dot{\vect{r}} =\bar{\vect{v}}_{\vect{k}} - \dot{\vect{k}} \times \bar{\vect{\Omega}}-\bar{\vect{\Omega}}_{\vect{k}t},
\quad \dot{\vect{k}} =-\vect{E}(t), \label{EOM}
\end{align}
where $\bar{\vect{v}}_{\vect{k}}=\vect{\nabla}_{\vect{k}} \bar{\epsilon}_k$ is the band velocity
that considers the field-induced correction,
$\bar{\vect{\Omega}}=\vect{\nabla}_{\vect{k}} \times \vect{\mathcal{\bar{A}}}$ is the Berry curvature
that also considers the field-induced correction,
and $\bar{\Omega}_{\vect{k}t}^\alpha=\partial_\alpha \mathcal{A}^t-\partial_t \bar{\mathcal{A}}^\alpha$ is
the Berry curvature defined in the $(\vect{k}, t)$ parameter space,
where $\mathcal{A}^t=\langle \bar{u} | i\partial_t|\bar{u}\rangle$ with $|\bar{u}\rangle$
the time-dependent Bloch state,
and $\bar{\mathcal{A}}^\alpha=\mathcal{A}^\alpha+\mathcal{A}^{\alpha,E}$ stands for
the Berry connection up to the first order of electric field.
By solving Eq.(\ref{EOM}), we find
\begin{align}
\dot{\vect{r}} = \bar{\vect{v}}_{\vect{k}} + \vect{E}(t) \times \bar{\vect{\Omega}}-\bar{\vect{\Omega}}_{\vect{k}t}.
\end{align}
Correspondingly, the intrinsic current density, defined as $\vect{J}=\int_k f \dot{\vect{r}}$, is given by
\begin{align}
\vect{J} = \int_k f \left[ \bar{\vect{v}}_{\vect{k}}+\vect{E}(t)\times\bar{\vect{\Omega}}-\bar{\vect{\Omega}}_{\vect{k}t} \right].
\label{current}
\end{align}
Furthermore, by restoring the band summation and focusing on the linear intrinsic current response,
from the last two terms of Eq.(\ref{current}) we obtain:
\begin{align}
J_\beta = \sum_{n} \int_k f_n \left[ \Omega^{\beta\alpha}_nE_\alpha(t)+G^{\beta\alpha}_n \partial_t E_\alpha(t) \right],
\label{semifinal}
\end{align}
which corresponds to the conduction current induced by Berry curvature and
the displacement current induced by quantum metric, respectively,
as given by Eq.(\ref{hall1}) and the adiabatic contribution of Eq.(\ref{Gab}) [that is Eq.(\ref{first})]
in quantum theory, respectively. Note that the first term $\bar{\vect{v}}$ of Eq.(\ref{current})
can not contribute a linear current due to 
$\bar{\epsilon}_n=\epsilon-1/2E_\alpha(t)G^{\alpha\beta}_nE_\beta$ \cite{XiaoCspin}.
In addition, for the last term of Eq.(\ref{current}),
we have used $\mathcal{A}^{\alpha, E}_n=G_n^{\beta\alpha}E_\alpha(t)$
for $\partial_t \bar{\mathcal{A}}^\alpha_n$
and ignored $\partial_\alpha \mathcal{A}^t_n$
since $\mathcal{A}^t_n \equiv \langle \bar{u}_n|i\partial_t|\bar{u}_n\rangle$
with $|\bar{u}_n \rangle = |u_n \rangle + \sum_{m \ne n} (\vect{E} \cdot \vect{r}_{mn}/\epsilon_{mn}) |u_m \rangle$
can not lead a contribution linear in the electric field.
Finally, we wish to remark that the semiclassical formulation is also valid for gapless systems
since the concept of electric polarization is not used in the derivation.
In fact, the static screening for metal in AC transport is not well defined \cite{screen}.


\section{The calculation details for Eq.(\ref{JxJy}), Eq.(\ref{JxJy1}), and Eq.(\ref{sigmaD})}

In this appendix, we present the calculation details for Eq.(\ref{JxJy}), Eq.(\ref{JxJy1}),
and Eq.(\ref{sigmaD}), given in the main text.

\subsection{Eq.(\ref{JxJy})} \label{AppC1}
For Dirac Hamiltonian Eq.(\ref{Ham1}),
the frequency-dependent integrand for the nonvanishing displacement current conductivity is reproduced as:
\begin{align}
\mathcal{G}_{\pm}^{xx}=\pm \dfrac{v_x^2(v_y^2k_y^2+\Delta^2)}{h^3(4h^2-\omega^2)},
\quad                                                                    
\mathcal{G}_{\pm}^{yy}=\pm \dfrac{v_y^2(v_x^2k_x^2+\Delta^2)}{h^3(4h^2-\omega^2)},
\end{align}
where $h^2=v_x^2k_x^2+v_y^2k_y^2+\beta^2$.
When the chemcial potential $\mu \in [-\Delta, \Delta]$, at zero temperature we find that
\begin{align}
J^D_{x}
&=
-E_x \omega \sin(\omega t) \int_k \mathcal{G}^{xx}_{-}
=
-E_x \omega \sin(\omega t) \iint \dfrac{dk_x dk_y}{4\pi^2}
\dfrac{v_x^2(v_y^2k_y^2+\Delta^2)}{(v_x^2k_x^2+v_y^2k_y^2+\Delta^2)^{3/2}\left[\omega^2-4(v_x^2k_x^2+v_y^2k_y^2+\Delta^2)\right]}
\nonumber \\
&=
-\dfrac{E_x \omega \sin(\omega t)v_x}{4\pi^2v_y} \iint d\bar{k}_x d\bar{k}_y
\dfrac{\bar{k}_y^2+\Delta^2}{(\bar{k}_x^2+\bar{k}_y^2+\Delta^2)^{3/2}\left[\omega^2-4(\bar{k}_x^2+\bar{k}_y^2+\Delta^2)\right]}
\end{align}
where we have set $\bar{k}_x=v_xk_x$ and $\bar{k}_y=v_yk_y$.
Let $(\bar{k}_x, \bar{k}_y)=\bar{k}(\cos\theta,\sin\theta)$, we find
\begin{align}
J^D_{x}
&=
-\dfrac{E_x \omega \sin(\omega t)v_x}{4\pi^2v_y} \int_0^{+\infty}\int_0^{2\pi}\bar{k} d\bar{k}d\theta 
\dfrac{\bar{k}^2\sin^2\theta+\Delta^2}{(\bar{k}^2+\Delta^2)^{3/2}\left[\omega^2-4(\bar{k}^2+\Delta^2)\right]}
\nonumber \\
&=
-\dfrac{E_x \omega \sin(\omega t)v_x}{4 \pi v_y} \int_0^{+\infty}\bar{k} d\bar{k}
\dfrac{\bar{k}^2+2\Delta^2}{(\bar{k}^2+\Delta^2)^{3/2}\left[\omega^2-4(\bar{k}^2+\Delta^2)\right]}
\end{align}
Furthermore, let $u^2=\bar{k}^2+\Delta^2$, we have $\bar{k}d\bar{k}=udu$ and find:
\begin{align}
J^D_{x}
&=
-\dfrac{E_x \omega \sin(\omega t) v_x}{4\pi v_y} 
\int_{\Delta}^{+\infty}
\dfrac{u^2+\Delta^2}{u^3 \left(\omega^2-4u^2\right)} udu
=
-\dfrac{E_x \omega \sin(\omega t) v_x}{4 \pi v_y} 
\left(
\int_{\Delta}^{+\infty}
\dfrac{1}{\omega^2-4u^2} du
+
\Delta^2\int_{\Delta}^{+\infty}
\dfrac{1}{u^2 \left(\omega^2-4u^2\right)} du
\right)
\nonumber \\
&=
-\dfrac{\Delta^2E_x \omega \sin(\omega t) v_x }{4\pi v_y} 
\int_{\Delta}^{+\infty}
\left[
\dfrac{1}{2\omega\Delta^2}
\left(
\dfrac{1}{\omega-2u}
+
\dfrac{1}{\omega+2u}
\right)
+
\dfrac{1}{\omega^2 u^2}
+
\dfrac{2}{\omega^3}
\left(
\dfrac{1}{\omega+2u}
+
\dfrac{1}{\omega-2u}
\right)
\right]
du
\nonumber \\
&=
-\dfrac{\Delta^2E_x \sin(\omega t) v_x}{4\pi v_y}
\left[
\dfrac{1}{2\Delta^2} \left( \dfrac{1}{2} \ln|\dfrac{2u+\omega}{2u-\omega}| \right)^{+\infty}_\Delta
+
\dfrac{1}{\omega} \left( -\dfrac{1}{u} \right)^{+\infty}_\Delta
+
\dfrac{2}{\omega^2} \left( \dfrac{1}{2} \ln|\dfrac{2u+\omega}{2u-\omega}| \right)^{+\infty}_\Delta
\right]
\nonumber \\
&=
-\dfrac{\Delta^2E_x \sin(\omega t) v_x}{4\pi v_y}
\left[
\dfrac{1}{\omega \Delta}
-
\left(
\dfrac{1}{4\Delta^2}
+
\dfrac{1}{\omega^2}
\right)
\ln\left| \dfrac{2\Delta+\omega}{2\Delta-\omega} \right|
\right],
\label{ingap}
\end{align}
Following a similar way, we have:
\begin{align}
J^D_{y}
&=
-\dfrac{\Delta^2E_y \sin(\omega t) v_y}{4\pi v_x}
\left[
\dfrac{1}{\omega \Delta}
-
\left(
\dfrac{1}{4\Delta^2}
+
\dfrac{1}{\omega^2}
\right)
\ln\left| \dfrac{2\Delta+\omega}{2\Delta-\omega} \right|
\right],
\label{ingap1}
\end{align}
In addition, when the chemical potential $\mu$ penetrates the lower band, we find:
\begin{align}
J^D_{x}
&=
-\dfrac{E_x \omega \sin(\omega t) v_x}{4 \pi v_y} 
\int_{\sqrt{\mu^2-\Delta^2}}^{+\infty}
\dfrac{\bar{k}^2+2\Delta^2 }{(\bar{k}^2+\Delta^2)^{3/2}\left[\omega^2-4(\bar{k}^2+\Delta^2)\right]}\bar{k}d\bar{k},
\\
J^D_{y}
&=
-\dfrac{E_y \omega \sin(\omega t) v_y}{4 \pi v_x} 
\int_{\sqrt{\mu^2-\Delta^2}}^{+\infty}
\dfrac{\bar{k}^2+2\Delta^2 }{(\bar{k}^2+\Delta^2)^{3/2}\left[\omega^2-4(\bar{k}^2+\Delta^2)\right]}\bar{k}d\bar{k},
\end{align}
which can be similarly calculated as:
\begin{align}
J^D_{x}
&=
-\dfrac{\Delta^2 E_x \sin(\omega t) v_x}{4\pi v_y}
\left[
\dfrac{1}{\omega |\mu|}
-
\left(
\dfrac{1}{4|\mu|^2}
+
\dfrac{1}{\omega^2}
\right)
\ln\left| \dfrac{2|\mu|+\omega}{2|\mu|-\omega} \right|
\right],
\\
J^D_{y}
&=
-\dfrac{\Delta^2 E_y \sin(\omega t) v_y}{4\pi v_x}
\left[
\dfrac{1}{\omega |\mu|}
-
\left(
\dfrac{1}{4|\mu|^2}
+
\dfrac{1}{\omega^2}
\right)
\ln\left| \dfrac{2|\mu|+\omega}{2|\mu|-\omega} \right|
\right].
\end{align}
This is true when the Fermi level penetrate the upper band,
where we need to sum over the bands.
By assuming that $\Delta \gg \hbar \omega$, we find
\begin{align}
\ln\left|\dfrac{1+\omega/2|\Delta|}{1-\omega/2|\Delta|}\right|
&=
\omega/|\Delta|+\omega^3/3|\Delta|^3+O(\omega^5/|\Delta|^5),
\\
\ln\left|\dfrac{1+\omega/2|\mu|}{1-\omega/2|\mu|}\right|
&=
\omega/|\mu|+\omega^3/3|\mu|^3+O(\omega^5/|\mu|^5),
\end{align}
therefore we finally obtain:
\begin{align}
J^D_{x}
=
\begin{cases}
\dfrac{e^2}{\hbar}
\dfrac{ 7 \hbar\omega E_x \sin(\omega t) v_x}{48\pi\Delta v_y}, \mu \in [-\Delta, \Delta] \\
\dfrac{e^2}{\hbar}
\dfrac{ 7 \hbar\omega \Delta^2 E_x \sin(\omega t) v_x}{48\pi |\mu|^3 v_y}, \mu \notin [-\Delta, \Delta]
\end{cases},
\quad
J^D_{y}
=
\begin{cases}
\dfrac{e^2}{\hbar}
\dfrac{ 7 \hbar\omega E_y \sin(\omega t) v_y}{48\pi\Delta v_x}, \mu \in [-\Delta, \Delta] \\
\dfrac{e^2}{\hbar}
\dfrac{ 7 \hbar\omega \Delta^2 E_y \sin(\omega t) v_y}{48\pi |\mu|^3 v_x}, \mu \notin [-\Delta, \Delta] 
\end{cases}
\end{align}
which gives the Eq.(\ref{JxJy}) by taking $v_x=v_y$ in the main text,
where $e$ and $\hbar$ are restored by dimension analysis.

\subsection{Eq.(\ref{JxJy1})}  \label{AppC2}
By introducing an additional term $sv_yk_x\sigma_x$ into Eq.(\ref{Ham1}) (that breaks mirror symmetry $\mathcal{M}_y$),
the off-diagonal integrand under the adiabatic limit is given by:
\begin{align}
G^{xy}_{\pm}
=
\mp \dfrac{v_y^2(v_x^2k_xk_y-\Delta^2)}{4(v_x^2k_x^2+v_y^2(k_x+k_y)^2+\Delta^2)^{5/2}}
=
\mp \dfrac{v_x^2(v_x^2k^2\sin\theta\cos\theta-\Delta^2)}{4\left[v_x^2k^2(\cos^2\theta+1+2\sin\theta\cos\theta)+\Delta^2\right]^{5/2}}
,
\label{Gxy}
\end{align}
where we have assumed $v_x=v_y$ and used the polar coordinate: $\vect{k}=k(\cos\theta,\sin\theta)$.
When the chemical potential $\mu$ is inside the gap, we find:
\begin{align}
\sigma_{xy}^{D}
&\equiv
\dfrac{J_x}{\omega E_y \sin \omega t} 
=
\dfrac{1}{4\pi^2} \int_0^{\infty} \int_0^{2\pi} 
\dfrac{v_x^2(v_x^2k^2\sin\theta\cos\theta-\Delta^2)}{4\left[v_x^2k^2(\cos^2\theta+1+2\sin\theta\cos\theta)+\Delta^2\right]^{5/2}}
kdkd\theta
\nonumber \\
&=
\dfrac{1}{8\pi^2} \int_0^{\infty} \int_0^{2\pi} 
\dfrac{v_x^2(v_x^2k^2\sin\theta\cos\theta-\Delta^2)}{4\left[v_x^2k^2(\cos^2\theta+1+2\sin\theta\cos\theta)+\Delta^2\right]^{5/2}}
dk^2d\theta.
\end{align}
Let $u=v_x^2k^2(\cos^2\theta+1+2\sin\theta\cos\theta)+\Delta^2$, we find
\begin{align}
k^2 =\dfrac{u-\Delta^2}{v_x^2(\cos^2\theta+1+2\sin\theta\cos\theta)},
\quad
dk^2=\dfrac{du}{v_x^2(\cos^2\theta+1+2\sin\theta\cos\theta)}.
\end{align}
Therefore, we have:
\begin{align}
\sigma_{xy}^{D}
&=
\dfrac{1}{32\pi^2} 
\left(\int_{\Delta^2}^{\infty} \int_0^{2\pi} 
\dfrac{v_x^4\sin\theta\cos\theta(u-\Delta^2)}{v_x^4(\cos^2\theta+1+2\sin\theta\cos\theta)^2 u^{5/2}} du d\theta
-
\int_{\Delta^2}^{\infty} \int_0^{2\pi} 
\dfrac{v_x^2\Delta^2}{v_x^2(\cos^2\theta+1+2\sin\theta\cos\theta)u^{5/2}} du d\theta
\right)
\nonumber \\
&=
-\dfrac{1}{12\pi\Delta} \label{Hall1app}
\end{align}
where we have used 
\begin{align}
\int_{\Delta^2}^\infty \dfrac{1}{u^{5/2}} du = \dfrac{2}{3\Delta^3},
\qquad
\int_{\Delta^2}^\infty \dfrac{1}{u^{3/2}} du = \dfrac{2}{\Delta},
\end{align}
and
\begin{align}
\int_0^{2\pi} \dfrac{\sin\theta\cos\theta}{(\cos^2\theta+1+2\sin\theta\cos\theta)^2} d\theta
=-\pi,
\qquad
\int_0^{2\pi} \dfrac{1}{(\cos^2\theta+1+2\sin\theta\cos\theta)} d\theta
=2\pi.
\end{align}
In a similar way, when the chemical potential $\mu$ penerates the bands, we find:
\begin{align}
\sigma^{D}_{xy}=-\dfrac{\Delta^2}{12|\mu|^3}. \label{Hall2app}
\end{align}
Eq.(\ref{Hall1app}) together with Eq.(\ref{Hall2app}) gives the Eq.(\ref{JxJy1}) in the main text.

\subsection{Eq.(\ref{sigmaD})} \label{AppC3}
Within the adiabatic limit, $\mathcal{G}^{ab}_{\pm}$ reduces to
the Berry connection polarizability tensor $G^{ab}_{\pm}$,
which for Eq.(\ref{HRD}) are given by:
\begin{align}
G^{xx}_{\pm}
&=
\pm \dfrac{\left( \lambda_D^2 - \lambda_R^2  \right)^2 k_y^2}
{4\left[ (\lambda_D k_y+\lambda_R k_x)^2 + (\lambda_D k_x +\lambda_R k_y)^2 \right]^{5/2}},
\\
G^{yy}_{\pm}
&=
\pm \dfrac{\left( \lambda_D^2 - \lambda_R^2  \right)^2 k_x^2}
{4\left[ (\lambda_D k_y+\lambda_R k_x)^2 + (\lambda_D k_x +\lambda_R k_y)^2 \right]^{5/2}},
\\
G^{yx}_{\pm}
&=
\mp \dfrac{\left( \lambda_D^2 - \lambda_R^2  \right)^2 k_x k_y}
{4\left[ (\lambda_D k_y+\lambda_R k_x)^2 + (\lambda_D k_x +\lambda_R k_y)^2 \right]^{5/2}}.
\end{align}
We first focus on the Hall component. Particularly,
when the chemical potential $\mu$ penetrates the lower band, at zero temperature we find
\begin{align}
J^D_y
&=
\omega E_x \sin(\omega t)
\int_k
\Theta(\mu-\epsilon_{-}) G^{xy}_-
=
\dfrac{\omega E_x \sin(\omega t)}{(2\pi)^2} 
\iint
\dfrac{ \Theta(\mu-\epsilon_{-})\left( \lambda_D^2 - \lambda_R^2  \right)^2 k_x k_y }
{4\left[ (\lambda_D k_y+\lambda_R k_x)^2 + (\lambda_D k_x +\lambda_R k_y)^2 \right]^{5/2}}
dk_x dk_y.
\label{sigmaDapp}
\end{align}
Let
\begin{align}
x = \lambda_D k_y + \lambda_R k_x,
\qquad
y = \lambda_D k_x + \lambda_R k_y,
\label{xy}
\end{align}
we find
\begin{align*}
k_x = \dfrac{ \lambda_R x - \lambda_D y}{\lambda_R^2-\lambda_D^2},
\qquad
k_y = \dfrac{ \lambda_D x - \lambda_R y}{\lambda_D^2-\lambda_R^2}.
\end{align*}
Therefore
\begin{align}
& (\lambda_R^2-\lambda_D^2)^2k_x^2 = \lambda_R^2 x^2+\lambda_D^2y^2-2\lambda_R\lambda_Dxy,
\quad
(\lambda_R^2-\lambda_D^2)^2k_y^2 = \lambda_D^2 x^2+\lambda_R^2y^2-2\lambda_R\lambda_Dxy,
\\
&(\lambda_R^2-\lambda_D^2)^2k_x k_y = (\lambda_R^2+\lambda_D^2)xy-\lambda_R\lambda_D(x^2+y^2),
\quad
\epsilon_{\pm} 
=
\pm
\sqrt{x^2+y^2}.
\label{kxky}
\end{align}
where we have droped the $E_0$ term in $\epsilon_{\pm}$ by focusing on the linear dispersion regime.
Correspondingly, the Jacobi determinant is given by
\begin{align}
J=\left| \dfrac{\partial(k_x,k_y)}{\partial(x,y)} \right|
=
\begin{pmatrix}
+\dfrac{\lambda_R}{\lambda_R^2-\lambda_D^2} & -\dfrac{\lambda_D}{\lambda_R^2-\lambda_D^2} \\
+\dfrac{\lambda_D}{\lambda_D^2-\lambda_R^2} & -\dfrac{\lambda_R}{\lambda_D^2-\lambda_R^2}
\end{pmatrix}
=
\dfrac{\lambda_R^2-\lambda_D^2}{(\lambda_R^2-\lambda_D^2)^2}
=
\dfrac{1}{\lambda_R^2-\lambda_D^2}.
\label{Jacobi}
\end{align}
Substituting Eqs. (\ref{xy}-\ref{Jacobi}) into Eq.(\ref{sigmaDapp}) we obtain:
\begin{align}
J_{y}^D 
&=
-\dfrac{\lambda_R \lambda_D \omega E_x \sin(\omega t)}{16\pi^2(\lambda_R^2-\lambda_D^2)}
\iint \Theta(\mu +\sqrt{x^2+y^2}) \dfrac{1}{(x^2+y^2)^{3/2}} dx dy
=
-\dfrac{\lambda_R \lambda_D  \omega E_x \sin(\omega t)}{16\pi^2(\lambda_R^2-\lambda_D^2)}
\int_0^{\infty} \int_0^{2\pi} \Theta(\mu+\rho)\dfrac{1}{\rho^3} \rho d\rho d\theta
\nonumber \\
&=
-\dfrac{\lambda_R \lambda_D  \omega E_x \sin(\omega t)}{8\pi(\lambda_R^2-\lambda_D^2)}
\int_{-\mu}^{\infty} \dfrac{1}{\rho^2} d\rho
=
-\dfrac{e^2}{\hbar}
\dfrac{\hbar \omega }{|\mu|} C_{yx}E_x \sin(\omega t),
\end{align}
as given by Eq.(\ref{sigmaD}) in the main text, where $C_{yx}=\lambda_R\lambda_D/[8\pi(\lambda_R^2-\lambda_D^2)]$ 
and we have restored the $e$ and $\hbar$ by dimension analysis.
Note that we obtain the same result when the chemical potential $\mu$ penetrates the upper band,
which can be derived as follows:
\begin{align}
J_{y}^D 
&=
-\dfrac{\lambda_R \lambda_D  \omega E_x \sin(\omega t)}{8\pi(\lambda_R^2-\lambda_D^2)}
\int_{0}^{\infty} \dfrac{1}{\rho^2} d\rho
+
\dfrac{\lambda_R \lambda_D  \omega E_x \sin(\omega t)}{8\pi(\lambda_R^2-\lambda_D^2)}
\int_{0}^{\mu} \dfrac{1}{\rho^2} d\rho
=
-\dfrac{e^2}{\hbar}
\dfrac{\hbar \omega }{|\mu|} C_{yx} E_x \sin(\omega t).
\end{align}
In a similar way, the longitudinal linear displacement current can be evaluated as
\begin{align}
J_{a}^D
=
\dfrac{e^2}{\hbar}
\dfrac{\hbar \omega }{|\mu|} C_{aa} E_a \sin(\omega t),
\end{align}
as also given by Eq.(\ref{sigmaD}) in the main text, 
where $C_{aa}=(\lambda_R^2+\lambda_D^2)/[16\pi(\lambda_R^2-\lambda_D^2)]$.

\twocolumngrid


\begin{thebibliography}{00}
\bibitem{geometry}
P. T\"orm\"a,
Phys. Rev. Lett. 131, 240001 (2023).
\bibitem{Nagaosa2017}
Y. Tokura, M. Kawasaki, and N. Nagaosa,
Nat. Phys. 13, 1056-1068 (2017).
\bibitem{JEmoore2017}
B. Keimer and J.E. Moore,
Nat. Phys. 13, 1045-1055 (2017).
\bibitem{BHYan2021}
J.W. Xiao and B.H. Yan,
Nat. Rev. Phys. 3, 283–297 (2021).
\bibitem{light1}
J.Y. Ahn, G.Y. Guo, N. Nagaosa, and A. Vishwanath,
Nat. Phys. 18, 290-295 (2022).
\bibitem{light2}
Q. Ma, R.K. Kumar, S.Y. Xu, F. H. L. Koppens, and J. C. W. Song,
Nat. Rev. Phys. 5, 170-184 (2023).
\bibitem{light3}
Q. Ma, A. G. Grushin, and K.S. Burch,
Nat. Mater. 20, 1601-1614 (2021).
\bibitem{symmetry}
Y. Tokura and N. Nagaosa,
Nat. Communi. 9, 3740 (2018).
\bibitem{Nagaosa2010}
N. Nagaosa, J. Sinova, S. Onoda, A. H. MacDonald, and N. P. Ong,
Rev. Mod. Phys. 82, 1539 (2010)
\bibitem{Niu2010}
D. Xiao, M.C. Chang, and Q. Niu,
Rev. Mod. Phys. 82, 1959 (2010).
\bibitem{FuBCD}
I. Sodemann and L. Fu,
Phys. Rev. Lett. 115, 216806 (2015).
\bibitem{MaQ2019}
Q. Ma, S.-Y. Xu, H. Shen, D. MacNeill, V. Fatemi, T.-R. Chang, A.M.M. Valdivia,
S. Wu, Z. Du, C.-H. Hsu, S. Fang, Q.D. Gibson, K. Watanabe, T. Taniguchi,
R.J. Cava, E. Kaxiras, H.-Z. Lu, H. Lin, L. Fu, N. Gedik, and P. Jarillo-Herrero,
Nature 565, 337 (2019).
\bibitem{KK2019}
K. Kang \textit{et al.},
Nat. Mater. 18, 324–328 (2019)
\bibitem{LuNRP2021}
Z. Z. Du, H.-Z. Lu, and X. C. Xie,
Nat. Rev. Phys. 3, 744 (2021).
\bibitem{QGT1}
G. P. Provost and G. Vallee,
Commut. Math. Phys. 76, 289 (1980).
\bibitem{QGT2}
J. Anandan and Y. Aharonov,
Phys. Rev. Lett. 65, 1697 (1990).
\bibitem{QGT3}
M. V. Berry,
Proc. R. Soc. London A 392, 45 (1984).
\bibitem{GaoY2014}
Y. Gao, S.A. Yang, and Q. Niu,
Phys. Rev. Lett. 112, 166601 (2014).
\bibitem{GaoY2021}
C. Wang, Y. Gao, and D. Xiao,
Phys. Rev. Lett. 127, 277201 (2021).
\bibitem{XiaoC2021}
H. Liu, J. Zhao, Y.-X. Huang, W. Wu, X.-L. Sheng, C. Xiao, and S. A. Yang,
Phys. Rev. Lett. 127, 277202 (2021).
\bibitem{XuSY2023}
A. Gao \textit{et al.},
Science 381, 181 (2023).
\bibitem{Gao2023}
N. Wang \textit{et al.}
Nature 621, 487 (2023).
\bibitem{CulcerQMD}
S. Lahiri, K. Das, R.B. Atencia, D. Culcer, and A. Agarwal,
arXiv:2207.02178 (2022).
\bibitem{ZGZhuQMD}
Y.D. Wang, Z.F. Zhang, Z.G. Zhu, and G. Su,
arXiv:2207.01182 (2022).
\bibitem{ZhangScience}
S. Murakami, N. Nagaosa, and S. C. Zhang,
Science 301, 1348 (2003).
\bibitem{Jackson}
J.D. Jackson, Classical Electrodynamics (third edition).
\bibitem{Marder}
M. P. Marder, Condensed Matter Physics (John Wiley \& Sons, Hoboken, 2010).
\bibitem{Nagaosa}
N. Nagaosa,
Annals of Physics 447, 169146 (2022).
\bibitem{KS-V1993}
R. D. King-Smith and D. Vanderbilt,
Phys. Rev. B 47, 1651 (1993).
\bibitem{V-KS1993}
D. Vanderbilt, R.D. King-Smith,
Phys. Rev. B 48, 4442 (1993).
\bibitem{Resta1994}
R. Resta, Rev. Mod. Phys. 66, 899 (1994).
\bibitem{XiaoCmagnetization}
C. Xiao, B.G. Xiong, and Q. Niu,
Phys. Rev. B 104, 064433 (2021).
\bibitem{Sipe1995} C. Aversa and J. E. Sipe, Phys. Rev. B 52, 14636 (1995).
\bibitem{Gonze2001} R. W. Nunes and X. Gonze, Phys. Rev. B 63, 155107 (2001).
\bibitem{Vanderbilt2002} I. Souza, J. Iniguez, and D. Vanderbilt, Phys. Rev. Lett. 89, 117602 (2002).
\bibitem{Wang2022}
C. Wang, R.-C. Xiao, H. Liu, Z. Zhang, S. Lai, C. Zhu, H. Cai,
N. Wang, S. Chen, Y. Deng, Z. Liu, S. A. Yang, and W.-B. Gao,
Natl. Sci. Rev. 9, nwac020 (2022).
\bibitem{YSA2021}
S. Lai, H. Liu, Z. Zhang, J. Zhao, X. Feng, N. Wang, C. Tang, Y. Liu, K. S. Novoselov, S. A. Yang, and W. B. Gao,
Nat. Nanotechnol. 16, 869 (2021).
\bibitem{XiaoCspin}
C. Xiao, W.K. Wu, H. Wang, Y.-X. Huang, X.L. Feng, H.Y. Liu, G.-Y. Guo, Q. Niu, and S. A. Yang,
Phys. Rev. Lett. 130, 166302 (2023).
\bibitem{screen}
K. Arakawa, T. Hayashida, K. Kimura, R. Misawa, T. Nagai, T. Miyamoto, H. Okamoto, F. Iga, and T. Kimura,
Phys. Rev. Lett. 131, 236702 (2023).
\bibitem{XiaoCdynamical}
C. Chen, D. Zhai, C. Xiao, W. Yao,
arXiv:2303.09973.
\bibitem{XiaoC2023}
D. Zhai, C. Chen, C. Xiao, and W. Yao,
Nat. Communi. 14 (1), 1961 (2023).
\bibitem{Sipe2000}
J. E. Sipe and A. I. Shkrebtii,
Phys. Rev. B 61, 5337 (2000).
\bibitem{JEMoore2010}
J. E. Moore and J. Orenstein,
Phys. Rev. Lett. 105, 026805 (2010).
\bibitem{JustingSong}
Q. Ma, R.K. Kumar, S.-Y. Xu, F.H.L. Koppens, and J.C.W. Song,
Nat. Rev. Phys. 5, 170-184 (2023).
\bibitem{Y-Gao1}
O. Bleu, G. Malpuech, Y. Gao, and D. D. Solnyshkov, Phys. Rev. Lett. 121, 020401 (2018).
\bibitem{Qiao}
Z. Liu, Z.H. Qiao, Y. Gao, Q. Niu, arXiv: 2212.08875.
\bibitem{Wanghua2020}
H. Wang and X.F. Qian, npj Comput. Mater. 6, 199 (2020).
\bibitem{anisotropy} R. E. Newnham,
Properties of Materials: Anisotropy, Symmetry, Structure
(Oxford University Press, 2005).
\bibitem{bilbao}
S. V. Gallego, J. Etxebarria, L. Elcoro, E. S. Tasci, and J.
M. Perez-Mato, Acta Crystallogr. Sect. A 75, 438 (2019).
\bibitem{KTlaw}
C.-P. Zhang, X.-J. Gao, Y.-M. Xie, H. C. Po, and K. T. Law,
Phys. Rev. B 107, 115142 (2023).
\bibitem{strain}
B. T. Zhou, C.P. Zhang, and K.T. Law,
Phys. Rev. Applied 13, 024053 (2020).
\bibitem{RD1}
J. Schliemann and D. Loss,
Phys. Rev. B 68, 165311 (2003).
\bibitem{RD2}
S.Q, Shen,
Phys. Rev. B 70, 081311(R) (2004).
\bibitem{RD3}
O. Pal and T.K. Ghosh,
Phys. Rev. B 109, 035202 (2024).
\bibitem{twist1}
E. Y. Andrei and A.H. MacDonald,
Nat. Mater. 19, 1265–1275 (2020).
\bibitem{twist2}
L. Balents, C.R. Dean, D.K. Efetov, and A.F. Young,
Nat. Phys. 16, 725–733 (2020).
\bibitem{twist3}
D.M. Kennes \textit{et al.},
Nat. Phys. 17, 155–163 (2021).
\bibitem{twist4}
E.Y. Andrei \textit{et al.},
Nat. Rev. Mater. 6, 201–206 (2021).
\bibitem{twist5}
C.N. Lau, M.W. Bockrath, K.F. Mak, and F. Zhang,
Nature 602, 41–50 (2022).
\bibitem{JAhn}
J. Ahn, G.Y. Guo, and N. Nagaosa,
Phys. Rev. X 10, 041041 (2020).
\bibitem{terahertz}
J. Liu, F. Xia, D. Xiao, J. G. de Abajo, and D. Sun,
Nat. Mater. 19, 830 (2020).
\bibitem{terahertz1}
Y. Zhang and L. Fu,
Proc. Natl. Acad. Sci., 118 21 (2021).
\end{thebibliography}
\end{document}